\documentclass[prb,aps,amsmath,amsfonts,amssymb,twocolumn,superscriptaddress,footinbib,showpacs, 10pt]{revtex4}
\usepackage{times,xspace}
\usepackage{amsfonts}
\usepackage{bm}
\usepackage{amssymb}
\usepackage[final]{graphicx}
\usepackage{color}

\begin{document}
 \title{Two Dimensional Electrons in (100)  Oriented Silicon
Field Effect Structures in the Region of Low Concentrations
and High Mobilities }

\author{V. T. Dolgopolov}
\affiliation{Institute of Solid State Physics, Russian Academy of Sciences,
Chernogolovka, Moscow region, 142432 Russia,
e-mail: dolgop@issp.ac.ru}

\begin{abstract}
A comparative analysis of experimental data on electron transport in Si (100) MOSFETs in the region of high
mobilities and strong electron-electron interaction is carried out. It is shown that electrons can be described
by the model of a noninteracting gas with the renormalized mass and Lande factor, which allows experimentally verifiable predictions.
\end{abstract}

\maketitle

Although the experimental studies of the two-
dimensional electron gas in silicon field-effect structures already have a history of more than half a century
\cite{ando}, they still go on (see, e.g.,\cite{sar,kunz,pud}). Interest in this
field of research is mainly related to the possible
metal-insulator transition \cite{kr1}, as well as the manifestations of strong electron-electron interaction \cite{kunz,pud,sar,kap,kl,kn1}
 for the lowest attainable electron densities on
 the order of $10^{11}$  $cm^{-2}$. In this range of electron densities, the effect of electron-electron interaction cannot
be described in terms of a perturbation theory series,
because the expansion parameter $r_s$ is considerably
larger than unity. Therefore, experimental results
should be interpreted very cautiously and, in the first
place, one needs to identify the common features
revealed in experiments on samples from different
manufacturers by different research groups using different methods of experimental data treatment. In this
publication, a part of this program is performed and
some further experiments are suggested.

Almost noninteracting quasiparticles
carrying a
unit charge exist only very close to the Fermi energy.
For them, one can define the Fermi momentum $p_F$,
determined by the number of free electrons introduced into the originally neutral  system; the velocity $v_F$; and the Lande factor $g^*$, which
characterizes only the quasiparticles at the Fermi surface. Each of these quantities can be measured experimentally.

The most convenient and accurate experimental
method to find the relation between the Fermi
momentum and Fermi velocity is to determine the
electron effective mass $m = p_F/v_F$ from the analysis of
the temperature dependence of the amplitude and
shape of the Shubnikov-de Haas oscillations. It is
appropriate to begin the analysis of experimental
results with measurements of this kind. However, it is
necessary to take into account a number of factors that
compel one to treat with some caution even these most
reliable data. (i) The amplitude of quantum oscillations needs to be sufficiently small both because of the
limitation on the validity of the Lifshitz-Kosevich
formula and because of the possible strong nonparabolicity of the spectrum \cite{khod} in the vicinity of the
Fermi level. The effective mass is overestimated owing
to the first factor and underestimated owing to the second. (ii) The quantum oscillation numbers have to be
sufficiently large to satisfy the assumptions made in
the Lifshitz-Kosevich theory. (iii) The quantum
relaxation time has to be independent of the temperature. Otherwise, the effective mass will be overestimated or underestimated, depending on the temperature range \cite{pud}.

Raw experimental data on the measurements of the
effective mass can be found in \cite{Gersh} and \cite{Rah}. These measurements were carried out by different experimental
groups and, furthermore, were performed on silicon
field-effect structures made by different manufacturers. According to \cite{pud,kl} , the experimental data
obtained for the effective mass can be well described by
the interpolation formula
\begin{equation}
m = 0.205m_e (1+0.035r_s+0.00016r_s^4)
\label{eq1}
\end{equation}
in the range $8 > r_s > 1$.5, where $r_s = 2.63 (10/n_s)^{1/2}$ and
the electron density is measured in units of $10^{11} cm^{-2}$.

\begin{figure}
\centerline{\scalebox{.45}{\includegraphics{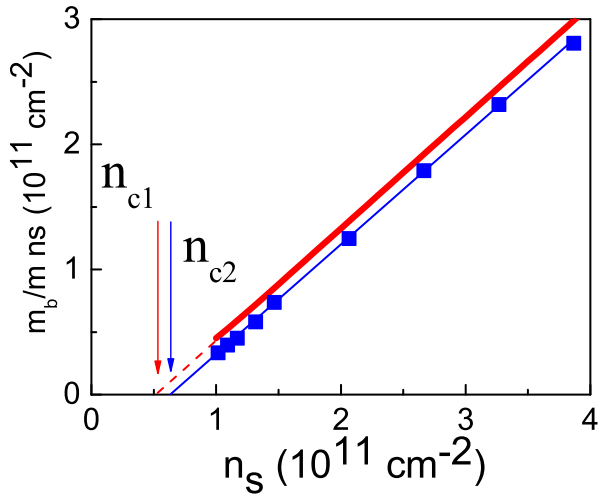}}} \caption{(Color online) Reciprocal effective mass versus the
electron density. The vertical axis is chosen so that the corresponding dependence for a system of noninteracting
electrons is a straight line with a slope of $45^o$ passing
through the origin. The thick solid line corresponds to the
experimental results fitted by Eq. (1). Symbols and the
straight line drawn through them represent the data from
[14]. Arrows show the critical densities arising upon the
linear extrapolation of the experimental data.} \label{fig1}
\end{figure}

Figure \ref{fig1}  shows the dependence of the effective
mass on the electron density as a plot of $m_b n_s/m$  versus
$n_s$, where $m_b = 0.19 m_e$ is the band effective mass. The
thick solid line corresponds to the effective mass given
by Eq.(\ref{eq1}), and points represent the experimental data
from \cite{Rah}. In the range of electron densities where
measurements were carried out, both sets of experimental data can be very well described by the dependence

\begin{equation}
m=m_b\frac{n_s}{(n_s-n_c)}.
\label{eq2}
\end{equation}

Indeed, in coordinates used in Fig.\ref{fig1} , each of these
data sets can be fitted by a straight line with a slope of
$42^o$. This corresponds to an experimentally obtained
mass $m_b$ that is $10 \%$ smaller than that known from \cite{ando}.
Equation (\ref{eq2}) describes data obtained for a large number of samples with different mobilities. It looks like
the critical density $n_c$ is determined by the interaction
between electrons and depends weakly on the chaotic
potential.

A behavior described by Eq.(\ref{eq2})  in the vicinity of
the quantum phase transition point in a strongly interacting two-dimensional electron system was predicted
in a number of publications \cite{dolg,dolg1,dobr,Am}. Note that Eq.(\ref{eq2})
remains valid up to electron densities as high as at least
$10^{12} cm^{-2}$.

The extrapolated values of the critical electron
density are somewhat different: $n_{c1}= 0.54*10^{11}cm^{-2}$
and $n_{c2} = 0.64*10^{11} cm^{-2}$. Most probably, the 15 $\%$
discrepancy occurs for purely technical reasons such
as different methods of experimental data treatment
and the overestimation of the experimental accuracy.
Certainly, the importance of more fundamental factors, for example, the different widths of the depletion
layer in samples supplied by different manufacturers
and, as a consequence, the difference in the Coulomb
interaction energy or the degree of disorder, cannot be
excluded.

Another reliably established fact is the independence of the effective mass of electrons in Si (100)
MOSFETs on the degree of spin polarization of the
electron system. This result was first obtained from the
analysis of the temperature dependence of the Shubnikov-de Haas oscillations in the presence of a magnetic field component parallel to the interface \cite{Rah}
and recently confirmed (at least, to the first approximation) by independent experiments \cite{pud} . This conclusion is also supported by some of the raw experimental
data presented in earlier publication \cite{Gersh} and by calculations for a multivalley electron system in the limiting
case of weak interaction \cite{masl}.
 
The further interpretation of the experimental data
requires additional assumptions that are, generally
speaking, poorly justified. Let us assume that a system
of strongly interacting particles can be described in
terms of a Fermi gas of noninteracting quasiparticles
with the same parameters as those at the Fermi level,
which depend only on the electron density. Indirectly,
this assumption is supported by a quantum Monte
Carlo calculation, which indicated that the difference
between the energies of a spin-polarized and a spin-unpolarized electron system is proportional to the
square of the degree of polarization, similarly to the
case of free electrons \cite{waintal}. In addition, let us extrapolate the linear dependence in Fig.\ref{fig1}  to the intersection
with the horizontal axis and further along this axis to
zero.

For $n_s>n_c$, the chemical potential level with
respect to the bottom of the band will be equal to    
\begin{equation}
\mu(n_s) = n_s\pi\hbar^2/2m = (n_s-n_c)\pi\hbar^2/2m_b
\label{eq3}
\end{equation}

According to this expression, the thermodynamic
density of states $\partial{n_s}/\partial{\mu}$ in this strongly interacting
electron system coincides with that in the original
noninteracting electron gas: $\partial{n_s}/\partial{\mu}= 2m_b/\pi\hbar^2$. The
validity of the latter relation can be tested experimentally. For this purpose, one needs to measure the difference between the MOSFET capacitances in zero
magnetic field and in those  quantizing magnetic fields
for which the thermodynamic density of states
becomes infinite (the method is described in detail,
e.g., in Ref. \cite{rem1}).

\begin{figure}
\centerline{\scalebox{.45}{\includegraphics{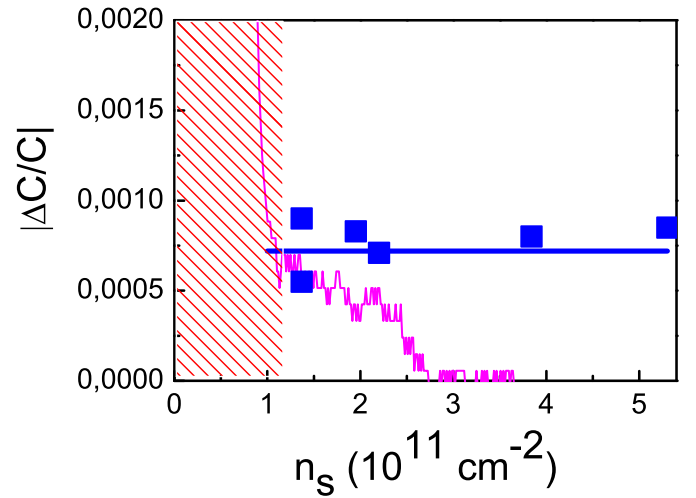}}} \caption{(Color online) Experimentally determined change
in the capacitance caused by the variation of the density of
states from the value corresponding to the spin unpolarized electron system to an infinitely large value (symbols)
and caused by the complete spin polarization of electrons
(noisy curve, B = 9.9 T). The solid horizontal line shows
the value expected for a gas of noninteracting electrons.
The shaded part of the plot at low electron densities corresponds to the region of electron localization in a magnetic
field completely spin polarizing the electron system.} \label{fig2}
\end{figure} 

The results of such measurements are shown in
Fig.\ref{fig2} . One can see that, in the absence of a magnetic
field, the thermodynamic density of states does coincide within the experimental error with that of noninteracting electrons, and the above assumptions lead to
conclusions that agree with the experiment.

Now, let us consider the case of a magnetic field
oriented parallel to the interface whose strength $B$ is
less than or equal to the field $B_p$ required to attain the
complete spin polarization. The magnetic moment $M$
per unit area of a two dimensional electron system is
determined by the density of states for a fixed number
of electrons and the difference between the energies of
electrons with the spin oriented along and opposite to
the field:

\begin{equation}
M = (\mu_B g)^2 B \frac{m_b}{2\pi \hbar^2}\frac{n_s}{n_s-n_c},
\label{eq4}
\end{equation}
where $\mu_B$ is the Bohr magneton and $g$ is the interaction renormalized Lande factor. It is assumed here, in
the context of experimental data testifying that the
effective mass is independent of the parallel component of the magnetic field, that the critical density $n_c$ is
also independent of the magnetic field.

Using the relation
\begin{equation}
\frac{\partial{M}}{\partial{n_s}} = - \frac{\partial{\mu}}{\partial{B}},
\label{eq5}
\end{equation}
the following expression for a quantity that can be
measured experimentally is obtained:
\begin{equation}
\frac{\partial{\mu}}{\partial{B}} = (\mu_B g)^2 B \frac{m_b}{2\pi \hbar^2}\frac{n_c}{(n_s-n_c)^2}.
\label{eq6}
\end{equation}
It is noteworthy that Eq.(\ref{eq6} contains only experimentally measurable quantities $g$ and $n_c$ and no fitting
parameters.

\begin{figure}
\centerline{\scalebox{.75}{\includegraphics{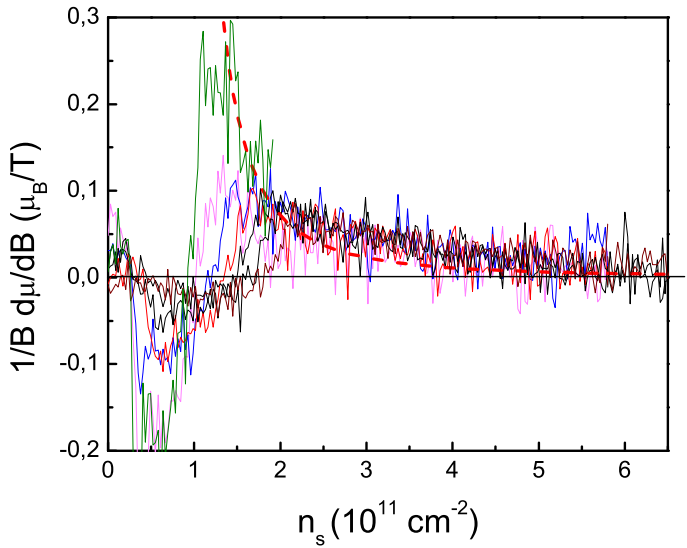}}} \caption{(Color online) Comparison of the experimental
results from [23] (for magnetic fields of 1.5, 2, 3, 4, 5, 6,
and 7 T) with (dashed line) the calculation according to
Eq.(\ref{eq6}).). } \label{fig3}
\end{figure} 

Experimental data on the derivative $\frac{\partial{\mu}}{\partial{B}}$ in different magnetic fields were obtained in Ref.s \cite{res,aniss} . In
Fig.\ref{fig3} , data from Ref.\cite{aniss} are compared with the curve calculated according to Eq.(\ref{eq6})  with \cite{skravc} $n_c = n_{c2}$ and $g=1.4g_0=2.8$.
One can see that (i) in full agreement
with Eq.9\ref{eq6}0, the experimental curves in the region
where they overlap do scale with the magnitude of the
magnetic field; (ii) the dependence given by Eq.(\ref{eq6})
qualitatively agrees with the experimental results; and
(iii) there is a noticeable quantitative disagreement.
Especially troublesome is the discrepancy in the density range $2*10^{11} cm^{-2}< n_s < 4*10^{11} cm^{-2}$, where it
is clearly larger than the possible error and cannot be
attributed to the finiteness of the temperature or the
inhomogeneity of the sample. Most probably, this discrepancy results from the use of the parameters
obtained for electrons at the chemical potential level
to describe electrons within the Fermi distribution.

\begin{figure}
\centerline{\scalebox{.55}{\includegraphics{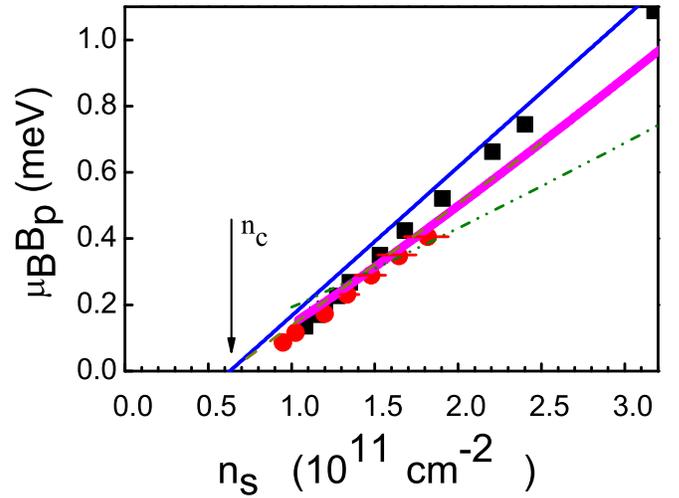}}} \caption{(Color online) Field corresponding to the complete
spin polarization of electrons versus their density. The thin
line shows the result of calculations using the fitting line
drawn through the experimental points in Fig.(\ref{fig1}) for $g =
1.4 g_0$; the thick line shows the expected field of complete
spin polarization calculated for the parameters at the
chemical potential level [4, 10]. Squares and circles correspond to the measurements of the magnetoresistance saturation field [24] and the results from [23], respectively. The
dash-dotted line is drawn according to the calculations of
[20] for an ideal electron system using the experimental
Lande factor.} \label{fig4}
\end{figure} 

Next, let us discuss the measurements of the field of
complete spin polarization $B_p$ for a magnetic field oriented parallel to the interface. The corresponding
experimental data are shown in Fig.(\ref{fig4}). For comparison, the results of independent measurements are
shown by lines. The thin solid line passing slightly
above the experimental points is obtained using the
interpolation of the data from Ref.\cite{Rah} and the Lande factor $g = 1.4 g_0$ according to the formula
 
\begin{equation}
\mu_B B_p = \frac{n_s \pi \hbar^2}{mg},
\label{eq7}
\end{equation}
which corresponds to the model of free electrons with
the renormalized mass and Lande factor. The thick
solid line is obtained from independent experiments
\cite{pud,kl}, where the product $gm$ is described by the
expression
\begin{equation}
gm = g_0m_b (1+0.21r_s+0.003r_s^3+0.0000045r_s^6).
\label{eq8}
\end{equation}

One can see that both the experimental points and the
two solid lines fall on straight lines with somewhat differing slopes at electron densities above $10^{11} cm^{-2}$. The
difference in the slopes does not exceed the systematic
experimental errors, and the extrapolation of both fitting lines to zero value of the field $B_p$ yields the same
critical density $n_c = 0.63*10^{11} cm^{-2}$. This value is
lower than that obtained by the extrapolation of the
straight line drawn through the experimental points,
which is probably caused by the abovementioned
overestimation of the energy required for the complete
spin polarization of the electron system.

The quantum Monte Carlo calculation of the complete spin polarization field \cite{waintal}  also yields a curve
close to a straight line in the region of interest on the
$(n_s, B_p)$ plane (see Fig.\ref{fig4}). However, its slope is noticeably smaller than the one obtained experimentally.
According to Ref.\cite{waintal} , the dependence of the complete spin polarization field $B_p$ on the electron density $n_s$ in
an electron system with disorder is qualitatively different: $B_p$ becomes zero for a finite value of $n_s$. This
behavior resembles that observed experimentally.
Nevertheless, even in this case, the calculation noticeably underestimates the energy required for complete
spin polarization for electron densities exceeding $2*10^{11} cm^{-2}$ . However, if the Lande factor determined for
electrons at the chemical potential level does not
describe electrons within the Fermi distribution and is
considered as an additional fitting parameter, a calculation \cite{waintal} using these two parameters (the degree of
disorder and the Lande factor $g$) can fit the experimental data in Fig.\ref{fig4} quite satisfactorily.

The transition of an electron system subjected to a
field $B_p$ to a completely spin polarized state must be
accompanied by an abrupt change in the thermodynamic density of states and, consequently, in the
capacitance of the MOSFET structure. An example of
a corresponding experimental plot is shown in Fig.\ref{fig2}.

Using Eq.(\ref{eq6}), it is easy to calculate the chemical
potential of a partially polarized electron system in a
magnetic field $B < B_p$:  

\begin{equation}
\mu(B,n_s)= \mu(0,n_n) +(\frac{\mu_BgB}{2})^2\frac{m_b}{\pi\hbar^2}\frac{n_c}{(n_s-n_c)^2},
\label{eq9}
\end{equation}
where $\mu(0,n_n)$ is given by Eq.(\ref{eq3}). Then, the inverse of
the thermodynamic density of states equals

\begin{equation}
\frac{\partial{\mu}}{\partial{n_s}} =\frac{\pi\hbar^2}{2m_b}- \frac{(\mu_BgB)^2m_bn_c} {2\pi\hbar^2(n_s-n_c)^3}.
\label{eq10}
\end{equation}
It varies most pronouncedly in the vicinity of the transition to the spin polarized state:
\begin{equation}
\frac{\partial{\mu}}{\partial{n_s}}(B_p) = \frac{\pi\hbar^2}{2m_b}(1-\frac{n_c}{(n_s(B_p)-n_c)}).
\label{eq11}
\end{equation}

The second term in the braces in Eq. (\ref{eq11}) is not small,
especially for $B_p \rightarrow 0$. Nevertheless, no experimental
manifestations of this term have been observed.
Indeed, according to Eq. (\ref{eq11}), one would expect the
occurrence of distinctly nonzero values of  $|\Delta C/C|$  for
densities above the observed jump in the capacitance.
This behavior is seen neither in Fig.\ref{fig2} nor in any of the
experimental curves in \cite{aniss}.

Such a significant discrepancy arises because the
quantity $\mu$ used in Eqs.(\ref{eq5},\ref{eq6},\ref{eq9},\ref{eq10},\ref{eq11}) is the
chemical potential measured with respect to the free electron level, while screening (and, thus, the capacitance) is determined by the difference between the
chemical potential and the energy at the bottom of the
two-dimensional electron subband. In experiments
where the shift of the subband bottom caused by external factors is not important, all above expressions are
valid. The second term in Eq. (\ref{eq9}) equals the shift of the
subband bottom, which can be easily seen by considering the limiting case $B \rightarrow B_p$. This second term has to
be omitted in the subsequent formulas that describe
screening. As a result, it should be expected that the
absolute value of the capacitance jump in the spin polarizing field will coincide with the horizontal line
in Fig.\ref{fig2}. Actually, according to the figure, this jump
proves to be 25$\%$ smaller.

Frequently, the two-dimensional electron system
in $Si MOSFETs$ is described in terms of another
approach based on the Landau theory of Fermi liquids. On one hand, this requires the introduction of a
larger number of parameters (amplitudes $F_0^s,  F_0^a, F_1^s$ 
instead of $m$ and $g$) to be determined from the
experiment; on the other hand, this makes it possible
to deal only with quasiparticles within a narrow energy
range in the vicinity of the chemical potential level.

Since the effective mass that determines the compressibility and thermodynamic density of states of an
electron system depends on two amplitudes $F_0^s$ and $ F_1^s$   as 
\begin{equation}
m_{com} =\frac{m}{1+F_0^s} = m_b\frac{1+ F_1^s}{1+F_0^s},
\label{eq12}
\end{equation}
the weak dependence of the capacitance jump
on the electron density (see Fig.\ref{fig2}) implies \cite{gold} that $F_0^s \simeq  F_1^s$.

It proves that the Fermi liquid parameter most sensitive to the interaction is $F_1^s$. According to Eqs.(\ref{eq2})
and (\ref{eq12}), $F_1^s$ can be expressed in the investigated range
of electron densities as
\begin{equation}
F_1^s = \frac{r_{sc}^2}{(r_{sc}^2-r_s^2)},
\label{eq13}
\end{equation}
and is larger than 2 for an electron density of $10^{11} cm^{-2}$,
which corresponds to $r_s \simeq 8$.

It is no surprise that the functional dependences of  $F_0^s$ 
and  $F_1^s$ are the same, because both amplitudes are
harmonics of the same function $F^{s}_{\bf k,\bf k'}$ , and the coincidence of their magnitudes within the experimental
error means that the dependence of $F^{s}_{\bf k,\bf k'}$ on the angle
between vectors ${\bf k}$  and ${\bf k'}$ is weak.

The parameter investigated in most detail (see \cite{kl})
is $F_0^a$, because it is expected that it can be determined
from two different experiments, i.e., from the measured values of the Lande factor and from the linear
temperature dependence of the elastic relaxation time \cite{zala}.
 The results obtained demonstrate within the
actual experimental accuracy that  $F_0^a$ depends weakly
on the electron density in the range from $1.9*10^{11} cm^{-2}$ to
about  $1*10^{12} cm^{-2}$.. At lower electron densities, a considerable increase in 
$|F_0^a|$ takes place, in fair agreement
with the result of Ref. \cite{gold}.

Thus, any one of the methods used for experimental data treatment leads to qualitatively coinciding
results for  $F_0^a$. One can hardly expect better agreement
because of both limited experimental accuracy and
limited applicability of the model used in Ref.\cite{zala} and
later publications. For example, the theory does not
take into account in any way that the scattering potential in Si MOSFETs changes radically in the range of
electron densities under consideration. For the lowest
densities, it is the screened Coulomb potential; as the
density increases, it changes over to interface roughness scattering. Possibly, the experimental results
obtained in Ref.\cite{vit}, which differ somewhat from those
considered above, follow from the change in the density
corresponding to the transition between different
scattering potentials.

In conclusion, let us once again stress the striking
fact that a system of strongly interacting two-dimensional electrons can be described with good accuracy
in terms of a gas of noninteracting electrons with the
renormalized mass and Lande factor, and this description demonstrates its predictive power. Hopes for further progress in the experimental investigation of the
two-dimensional electron system in silicon MOSFETs
in the region of low densities and high mobilities are
associated with precision measurements of the thermodynamic density of states.

I am grateful to S.V. Kravchenko, V.M. Pudalov,
and A.A. Shashkin for useful discussions. This study
was supported by the Russian Foundation for Basic
Research (project nos. 13-02-00095 and 13-02-
12127) and the Russian Academy of Sciences.


\begin{thebibliography}{110}
\bibitem{rem} To avoid misunderstanding, the expansion in the Landau amplitudes for the two-dimensional case used below has the form  $ F^{s,a}_{\bf k,\bf k'}= F^{s,a}_0 +2 \Sigma_{n=1}^\infty F^{s,a}_n Cos(n[\theta-\theta^{'}]) $.
\end{thebibliography}
 \end{document}